\documentclass[printer,onecolumn,traditabstract]{aa}
\usepackage{natbib}
\bibpunct{(}{)}{;}{a}{}{,}

\usepackage[pdftex]{graphicx}

\usepackage{setspace,ulem}
\usepackage{psfig,epsfig,amssymb,alltt}
\usepackage{latexsym}
\usepackage{indentfirst}
\usepackage[english]{babel}
\usepackage{graphicx}
\usepackage{epstopdf}
 
\usepackage{chemarrow}
\usepackage{psfig}
\usepackage{vmargin}
\usepackage{fancybox}
\usepackage{fancyheadings}
\usepackage{algorithmic}
\usepackage{algorithm}
\usepackage{theorem}
\usepackage{amssymb,amsmath}
 \usepackage{longtable,multirow,enumerate,threeparttable}
 \usepackage{amsfonts,upgreek,bbm,mathrsfs,mathabx}

\usepackage{url}
\usepackage[applemac]{inputenc}
\usepackage[T1]{fontenc}
\vspace{1cm}

\newcommand{\parenth}[1]{\left({#1}\right)}

\psfigurepath{./ps}

\title{Multichannel Poisson denoising and deconvolution on the sphere : Application to the Fermi Gamma Ray Space Telescope}
\author{J. Schmitt \inst{1}
\and J.L. Starck \inst{1}
\and J.M. Casandjian \inst{1}
\and J. Fadili \inst{2}
\and I. Grenier \inst{1}
}

\authorrunning{Schmitt et {\textit{al}}.}
\titlerunning{Poisson Deconvolution on the Sphere}

\institute{
\inst{1} CEA, Laboratoire AIM, CEA/DSM-CNRS-Universit\'e Paris Diderot,   \\
CEA, IRFU, Service d'Astrophysique,  Centre de Saclay,  F-91191 Gif-Sur-Yvette cedex, France \\
\inst{2} GREYC CNRS-ENSICAEN-Universit\'e de Caen, 6, Bd du Mar\'echal Juin, 14050 Caen Cedex, France}

\begin{document}
\date{\today}

\abstract{A multiscale representation-based denoising method for spherical data contaminated with Poisson noise, the multiscale variance stabilizing transform on the sphere (MS-VSTS), has been previously proposed. 
This paper first extends this MS-VSTS to spherical two and one dimensions data (2D-1D), where the two first dimensions are longitude and latitude, and the third dimension is a meaningful physical index such as energy or time.
We then introduce a novel multichannel deconvolution built upon the 2D-1D MS-VSTS, which allows us to get rid of both the noise and the blur introduced by the point spread function (PSF) in each energy (or time) band.
The method is applied to simulated data from the Large Area Telescope (LAT), the main instrument of the Fermi Gamma-Ray Space Telescope, which detects high energy gamma-rays in a very wide energy range (from 20 MeV to more than 300 GeV), and whose PSF is strongly energy-dependent (from about $3.5°$ at $100$ MeV to less than $0.1°$ at $10$ GeV).
%The simulated data set is composed of Fermi skymaps, corresponding to energy bands between $50$ MeV and $1.58$ GeV. Each skymap is a HEALPix map with $\mathrm{nside}=256$.
}

\maketitle


%\tableofcontents

\section{Introduction}
\subsection{Literature overview}
The gamma-ray sky has been studied with unprecedented sensitivity and image capability thanks to the Large Area Telescope (LAT), the main instrument of the Fermi Gamma-Ray Space Telescope \citep{Fermi}, in an energy range between 20 MeV to greater than 300 GeV. The detection of gamma-ray point sources is difficult for two main reasons : the Poisson noise and the instrument's point spread function (PSF). The Poisson noise is due to the fluctuations in the number of detected photons. Moreover, the effect of Poisson noise is strong because of the weakness of the fluxes of celestial gamma rays, especially outside the Galactic plane and far away from intense sources. The PSF width is strongly energy-dependent, varying from about $3.5°$ at $100$ MeV to less than $0.1°$ ($68\%$ containment) at $10$ GeV. Owing to large-angle multiple scattering in the tracker, the PSF has broad tails, for which the $95\%/68\%$ containment ratio may be as large as 3.

An extensive literature exists on Poisson noise removal and the interested reader may refer to \citet{schmitt_chapt11} and \citet{starck:book10} for a thorough review. Motivated by new X-ray and gamma-ray data challenges, several restoration methods have been released in astrophysics that are based on wavelets \citep{movit2009,Starck09:fermi3d,fay2011,Schmitt} or the Bayesian machinery \citep{conrad2007,norris2010}. 
Wavelets have also been used for source detection in Fermi data \citep{fermicat2010}, and a first Poisson denoising algorithm for spherical data was proposed in \citet{Schmitt}.
\citet{Starck09:fermi3d} developed a denoising approach that effectively handles multichannel data acquired on a cartesian grid, and where the third dimension can be any physically meaningful index such as wavelength, energy, or time. While traditional techniques integrate over all the third dimension in order to improve the signal-to-noise ratio (S/N) of the sources, their approach allows us to detect the sources while preserving all of their spectral information without sacrificing any the sensitivity. Nonetheless, none of these methods take into account the point spread function (PSF) of the instrument.
Deconvolution can be very helpful in many situations such as source identification or flux estimation in the low energy bands where the resolution degrades severely.
% , since  there is a the strong energy dependence of the LAT's PSF, and the recovering of the spectral information of point sources.
To the best of our knowledge, no general method for both multichannel denoising and deconvolving on the sphere has been developed in the literature.

\subsection{Contributions}
We propose a general framework for denoising and deconvolution that complies with all the above requirements, namely deals with \begin{enumerate}
  \item multichannel  data,
  \item acquired on the sphere,
  \item and contaminated by Poisson noise.
\end{enumerate}
Our approach builds upon the concept of {\textit{variance stabilization}} applied to the spherical wavelet transform coefficients \citep{starck2006}. This approach gives a multiscale representation of the Poisson data with variance-stabilized coefficients, which can be treated as if they were contaminated by a zero-mean white Gaussian noise. The developed algorithms are validated on simulated Fermi HEALPix multichannel cubes ($\mathrm{nside}=256$) with energy bands ranging from 50 MeV to 50 GeV.

\subsection{Paper organization}
The rest of the paper is organized as follows. In section~\ref{mrs2d1d}, we present a new wavelet transform for multichannel data on the sphere, and we show how the variance stabilization transform can be introduced into the 
decomposition. Section~\ref{fil_mrs2d1d} details the way that this transform can be used for Gaussian and Poisson noise removal. Section~\ref{dec_mrs2d1d} describes our deconvolution algorithm on the sphere for both mono-channel and multichannel spherical data. In Section~\ref{sec:conclusion}, we finally draw some conclusions and give possible perspectives of this work.

\subsection{Notations}
\label{subsec:notations}
For a real discrete-time filter whose impulse response is $h[i]$, $\bar{h}[i]=h[-i], ~ i \in \mathbb{Z}$ is its time-reversed version. The discrete circular convolution product of two signals is written $\star$, where the term circular represents periodic boundary conditions. The symbol $\delta[i]$ is the Kronecker delta. 

For the wavelet representation, the low-pass analysis filter is denoted $h$ and the high-pass is taken as $g=\delta - h$ throughout the paper. We denote the up-sampled version of $h$ as $h^{\uparrow j}[l] = h[l]$ if $l / 2^j \in \mathbb{Z}$ and $0$ otherwise. We define $h^{(j)} = \bar{h}^{\uparrow j-1}\star\cdots\star\bar{h}^{\uparrow 1}\star \bar{h}$ for $j\geqslant 1$ and $h^{(0)} = \delta$.

The scaling and wavelet functions {used} for the analysis (respectively, synthesis) are denoted $\phi$ (with $\phi(\frac{x}{2}) = \sum_k h[k] \phi(x-k), x \in {\mathbb{R}} ~ and ~ k \in {\mathbb{Z}}$) and $\psi$ (with $\psi(\frac{x}{2}) = \sum_k g[k] \phi(x-k), x \in {\mathbb{R}} ~ and ~ k \in {\mathbb{Z}}$) (respectively, $\widetilde{\phi}$ and $\widetilde{\psi}$). 

\section{Multiscale representation for multichannel spherical data with poisson noise}
\label{mrs2d1d}
\subsection{Fast undecimated 2D-1D wavelet decomposition/reconstruction on the sphere}
\label{subsec:fastmrs2d1d}

Our goal is to analyze multichannel data acquired on a sphere with a non-isotropic two-and-one-dimensional (2D-1D) wavelet, where the two first dimensions are spatial (longitude and latitude) and the third dimension is either the time or the energy. Since the dimensions do not have the same physical meaning, it appears natural that the wavelet scale along the third dimension (energy or time) should not be connected to the spatial scale. Hence, we define the wavelet function as
\begin{equation}
\psi( k_{\theta},k_{\varphi}, k_{t}) = \psi^{(\theta \phi)} (k_{\theta}, k_{\varphi}) \psi^{(t)} (k_t) ,
\end{equation}
where $\psi^{(\theta \phi)}$ is the spherical two dimensional (2D) spatial wavelet and $\psi^{(t)}$ the one-dimensional (1D) wavelet along the third dimension. Similarly to \citet{Starck09:fermi3d}, we consider only isotropic and dyadic spatial scales.\textbf{ We} 
build the discrete 2D-1D wavelet decomposition by first taking a spherical 2D undecimated wavelet transform for each $k_{t}$, followed by a 1D wavelet transform for each spatial wavelet coefficient along the third dimension.
  
% We propose a denoising method for 2D - 1D data on the sphere, where the two first dimensions are spatial (longitude and latitude) and the third dimension is either the time or the energy. We need to analyze the data with a non-isotropic wavelet, where the time or energy scale is not connected to the spatial scale. We use a spherical extension of the Fast Undecimated 2D-1D Wavelet Transform proposed in \citet{Starck09:fermi3d}.

%In order to have a fast algorithm for discrete data, we use wavelet functions associated to filter banks. 

Hence, for a given multichannel data set on the sphere $Y[k_{\theta},k_{\varphi},k_t]$ and after applying first the 2D spherical undecimated wavelet transform, we have the reconstruction formula
%Hence, our wavelet decomposition consists in applying first a IUWT on the sphere for each frame $k_z$. Using the spherical IUWT, we have the reconstruction formula:
\begin{equation}
\label{reconstiuwt}
Y[k_{\theta},k_{\varphi},k_{t}]  = a_{J_1}[k_{\theta},k_{\varphi},k_{t}] + \sum_{j_1=1}^{J1}w_{j_1}[k_{\theta},k_{\varphi},k_{t}], \qquad \forall k_{t} ,
\end{equation}
where $J_1$ is the number of spatial scales, $a_{J_1}$ is the (spatial) approximation subband, and $\{w_{j_1}\}_{j_1=1}^{J_1}$ are the (spatial) detail subbands. To simplify the notations in the sequel, we replace the two spatial indices by a single index $k_r$, which corresponds to the pixel index. Equation \eqref{reconstiuwt} now reads
\begin{equation}
\label{reconstiuwtpix}
Y[k_{r},k_{t}]  = a_{J_1}[k_{r},k_t] + \sum_{j_1=1}^{J1}w_{j_1}[k_{r},k_{t}], \qquad \forall k_{t} .
\end{equation}
For each spatial location $k_r$ and each 2D wavelet scale $j_1$, we then apply a 1D wavelet transform along $t$ on the spatial wavelet coefficients $w_{j_1}[k_r,\cdot]$ such that
\begin{equation}
\label{decompwavj1}
w_{j_1}[k_r,k_t] = w_{j_1,J_2} [k_r,k_t] + \sum_{j_2=1}^{J_2}w_{j_1,j_2}[k_r,k_t], \quad \forall(k_r,k_t),
\end{equation}
where $J_2$ is the number of scales along $t$. The approximation spatial subband $a_{J_1}$ is processed in a similar way, hence yielding
\begin{equation}
\label{decompaj}
a_{J_1}[k_r,k_t] = a_{J_1,J_2} [k_r,k_t] + \sum_{j_2=1}^{J_2} w_{J_1,j_2}[k_r,k_t], \qquad \forall(k_r,k_t) .
\end{equation}
Inserting Eqs. \eqref{decompwavj1} and \eqref{decompaj} into \eqref{reconstiuwtpix}, we obtain the 2D-1D spherical undecimated wavelet representation of $Y$:
\begin{equation}
\label{decomp2d1d}
Y[k_r,k_t] = a_{J_1,J_2}[k_r,k_t] + \sum_{j_1=1}^{J_1}w_{j_1,J_2} [k_r,k_t] + \sum_{j_2=1}^{J_2} w_{J_1,j_2} [k_r,k_t] + \sum_{j_1=1}^{J_1}\sum_{j_2=1}^{J_2} w_{j_1,j_2} [k_r,k_t] ~.
\end{equation}

\noindent
In this expression, four kinds of coefficients can be distinguished:
\begin{itemize}
  \item Detail-detail coefficients ($j_1 \leqslant J_1$ and $j_2 \leqslant J_2$):
  \begin{equation}
  \label{detaildetail}
w_{j_1,j_2}[k_r,\cdot] = (\delta - \bar{h}_{\mathrm{1D}}) \star (h_{\mathrm{1D}}^{(j_2-1)} \star a_{j_1-1}[k_r,\cdot]
- h_{\mathrm{1D}}^{(j_2-1)} \star a_{j_1}[k_r,\cdot]) .
  \end{equation}
  \item Approximation-detail coefficients ($j_1 = J_1$ and $j_2 \leqslant J_2$):
\begin{equation}
\label{approxdetail}
w_{J_1,j_2}[k_r,\cdot] = h_{\mathrm{1D}}^{(j_2-1)} \star a_{J_1}[k_r,\cdot] - h_{\mathrm{1D}}^{(j_2)} \star a_{J_1}[k_r,\cdot] .
\end{equation}
  \item Detail-approximation coefficients ($j_1 \leqslant J_1$ and $j_2 = J_2$):
\begin{equation}
\label{detailapprox}
w_{j_1,J_2}[k_r,\cdot] = h_{\mathrm{1D}}^{(J_2)} \star a_{j_1-1}[k_r,\cdot] - h_{\mathrm{1D}}^{(J_2)} \star a_{j_1}[k_r,\cdot] .
\end{equation}
  \item Approximation-approximation coefficients ($j_1 = J_1$ and $j_2 = J_2$):
\begin{equation}
\label{approxapprox}
a_{J_1,J_2}[k_r,\cdot] = h_{\mathrm{1D}}^{(J_2)} \star a_{J_1}[k_r,\cdot] .
\end{equation}
\end{itemize}

%\subsection{Poisson Noise}

\subsection{Multi-scale variance stabilizing transform on the sphere (MS-VSTS)}

\citet{Schmitt} proposed a multiscale variance stabilizing transform adapted for Poisson spherical data. This transform was dubbed the multi-scale variance stabilizing transform on the Sphere (MS-VSTS). The MS-VSTS is a multi-scale decomposition method designed for Poisson noise that is based on a variance stabilizing transform(VST). Poisson noise is indeed signal-dependent, which complicates its removal. The aim of a VST is to get rid of this signal-dependence by transforming a Poisson distribution into a Gaussian distribution of known variance. In a nutshell, the MS-VSTS consists of plugging a VST \textbf{into} a multi-scale transform--the isotropic undecimated wavelet transform on the sphere (IUWTS)-- in order to realize (approximately) Gaussian zero-mean multiscale coefficients with constant variance. The noise on coefficients can then be easily removed using Gaussian denoising methods such as wavelet shrinkage, as we see in the next section.

The MS-VSTS scheme is defined recursively by inserting a (nonlinear) square-root VST into the IUWTS steps, that is
%This section describes the MS-VSTS + IUWT, which is a combination of a square-root VST with the IUWT. The recursive scheme is:
\begin{equation}
\label{eq27}
\text{IUWTS}
\left\{\begin{array}{ccc}
a_j  & = &  h_{j-1} \star a_{j-1}  \\
d_j  & = & a_{j-1}  - a_j  
\end{array}\right.
 \Longrightarrow  
 \begin{array}{c}\text{MS-VSTS} \\  \text{(VST + IUWTS)} \end{array}
\left\{\begin{array}{ccc}
a_j  & = &  h_{j-1} \star a_{j-1} \\
d_j  & = & T_{j-1}(a_{j-1}) - T_j(a_j) 
\end{array}\right.,
\end{equation}
where $T_j$ is the VST operator on scale $j$
\begin{equation}
\label{eq28}
T_j(a_j) = b^{(j)} \mathrm{sign}(a_j+c^{(j)})\sqrt{|a_j + c^{(j)}|} ,
\end{equation}
with the VST constants $b^{(j)}$ and $c^{(j)}$ that depend solely on the filter $h$ and the scale level $j$. \citet{Zhang} showed that the MS-VSTS detail coefficients $d_j$ on locally homogeneous parts of the underlying intensity signal follow asymptotically a zero-mean normal distribution with an intensity-independent variance that relies only on the filter $h$ and the current scale $j$. Consequently, for a given $h$, both the stabilized variances and the constants $b^{(j)}$ and $c^{(j)}$ can be pre-computed once and for all \citep{Schmitt}.

\subsection{Multichannel MS-VSTS}

%To perform a Poisson denoising, we have to plug the MS-VST into the spherical 2D-1D undecimated wavelet transform.

We now extend the MS-VSTS machinery to the multichannel case. This amounts to plugging the VST into the spherical 2D-1D undecimated wavelet transform introduced in Section~\ref{subsec:fastmrs2d1d}. This again gives rise to four types of coefficients that take the following forms:
\begin{itemize}
  \item Detail-detail coefficients ($j_1 \leqslant J_1$ and $j_2 \leqslant J_2$):
  \begin{equation}
  \label{detaildetailmsvst}
w_{j_1,j_2}[k_r,\cdot] = (\delta - \bar{h}_{\mathrm{1D}}) \star \parenth{ T_{j_1-1,j_2-1} \parenth{{h}_{\mathrm{1D}}^{(j_2-1)} \star a_{j_1-1}[k_r,\cdot] } 
- T_{j_1,j_2-1}\parenth{h_{\mathrm{1D}}^{(j_2-1)} \star a_{j_1}[k_r,\cdot]}} .
  \end{equation}
  \item Approximation-detail coefficients ($j_1 = J_1$ and $j_2 \leqslant J_2$):
\begin{equation}
\label{approxdetailmsvst}
w_{J_1,j_2}[k_r,\cdot] = T_{J_1,j_2-1}\parenth{h_{\mathrm{1D}}^{(j_2-1)} \star a_{J_1}[k_r,\cdot]} - T_{J_1,j_2}\parenth{h_{\mathrm{1D}}^{(j_2)} \star a_{J_1}[k_r,\cdot]} .
\end{equation}
  \item Detail-approximation coefficients ($j_1 \leqslant J_1$ and $j_2 = J_2$):
\begin{equation}
\label{detailapproxmsvst}
w_{j_1,J_2}[k_r,\cdot] = T_{j_1-1,J_2} \parenth{h_{\mathrm{1D}}^{(J_2)} \star a_{j_1-1}[k_r,\cdot]} - T_{j_1,J_2} \parenth{h_{\mathrm{1D}}^{(J_2)} \star a_{j_1}[k_r,\cdot]} .
\end{equation}
  \item Approximation-approximation coefficients ($j_1 = J_1$ and $j_2 = J_2$):
\begin{equation}
\label{approxapproxmsvst}
a_{J_1,J_2}[k_r,\cdot] = h_{\mathrm{1D}}^{(J_2)} \star a_{J_1}[k_r,\cdot] .
\end{equation}
\end{itemize}

In summary, all 2D-1D wavelet coefficients $\{w_{j_1,j_2}\}_{j_1 \leqslant J_1,j_2 \leqslant J_2}$ are now stabilized, and the noise in all these wavelet coefficients is a zero-mean Gaussian with known variance that depends solely on $h$ on the resolution levels $(j_1,j_2)$. As before, these variances can be easily tabulated. %Denoising is however not straightforward because there is no explicit reconstruction formula available because of the form of the stabilization equations above. Formally, the stabilizing operators $T_{j_1,j_2}$ and the convolution operators along the spatial and temporal dimensions do not commute, even though the filter bank satisfies the exact reconstruction formula. To circumvent this difficulty, we propose to solve this reconstruction problem by using an iterative reconstruction scheme.

%exemple : transfo d'un dirac

\section{Application to multichannel denoising}
\label{fil_mrs2d1d}

We define $X$ to be the noiseless data and $Y$ their observed noisy version. In the case of the additive zero-mean white Gaussian noise, we have $Y \sim \mathcal{N}(X,\sigma^2)$, and for the Poisson noise $Y \sim \mathcal{P}(X)$. The main objective behind denoising is to estimate $X$ from $Y$.

\subsection{Warm-up: Gaussian noise}

We start with the simple and instructive case where the noise in $Y$ is additive white Gaussian. As the spherical 2D-1D undecimated wavelet transform described in Section~\ref{subsec:fastmrs2d1d} is linear, the noise remains Gaussian in the transform domain. Therefore, the thresholding strategies that have been developed for wavelet Gaussian denoising can be applied to the spherical 2D-1D wavelet transform. Denoting the thresholding operator as TH, the denoised 2D-1D estimate of $X$ obtained by thresholding the wavelet coefficients in Eq. \eqref{decomp2d1d} reads
\begin{equation}
\label{denoisegaussmc}
\widetilde{X}[k_r,k_t] = a_{J_1,J_2}[k_r,k_t] + \sum_{j_1=1}^{J_1}\text{TH}(w_{j_1,J_2} [k_r,k_t]) + \sum_{j_2=1}^{J_2} \text{TH}(w_{J_1,j_2} [k_r,k_t]) + \sum_{j_1=1}^{J_1}\sum_{j_2=1}^{J_2} \text{TH}(w_{j_1,j_2} [k_r,k_t]) .
\end{equation}
A typical choice of TH is the hard thresholding operator parametrized by the scalar threshold $\tau \geqslant 0$, i.e.
\begin{equation}
\label{hardthresh}
\text{TH}(x) = 
\begin{cases}
0 & \text{if } |x| < \tau, \\
x & \text{otherwise}. 
\end{cases}
\end{equation}
The threshold $\tau$ is typically chosen to be between three and five times the noise standard deviation.

\subsection{Poisson noise}

%Hence, all 2D-1D wavelet coefficients $w_{j_1,j_2}$ are now stabilized, and the noise on all these wavelet coefficients is Gaussian with known scale-dependent variance that depends solely on $h$. Denoising is however not straightforward because there is no explicit reconstruction formula available because of the form of the stabilization equations above. Formally, the stabilizing operators $T_{j_1,j_2}$ and the convolution operators along the spatial and temporal dimensions do not commute, even though the filter bank satisfies the exact reconstruction formula. To circumvent this difficulty, we propose to solve this reconstruction problem by using an iterative reconstruction scheme.

The treatment of Poisson noise is far more complicated than its Gaussian counterpart, the main reason being that the noise variance is equal to its mean. This is where the MS-VSTS comes into play. The role of the MS-VSTS is indeed to get rid of this dependence of the variance on the mean by ensuring that the transformed coefficients are
Gaussian with constant variance (without loss of generality, this variance can be assumed to be equal to $1$). In other words, after the MS-VSTS, we are brought to a Gaussian denoising problem where standard thresholding approaches apply. 

Nevertheless, denoising is not straightforward because there is no explicit reconstruction formula available because of the form of the non-linear stabilization equations above. Formally, the stabilizing operators $T_{j_1,j_2}$ and the convolution operators along the spatial and the third dimensions do not commute, even though the filter bank satisfies the exact reconstruction formula. To circumvent this difficulty, we propose to solve this reconstruction problem by advocating an iterative reconstruction scheme.

\subsection{Iterative reconstruction}
We define $\mathbf{W}$ to be the transform operator associated with the 2D-1D IUWTS  described in Section~\ref{subsec:fastmrs2d1d}, and $\mathbf{R}$ to be its inverse transform. We define $\mathcal{M}$ to be the multiresolution support, which is determined by the set of significant coefficients detected among $\mathbf{W}Y$ at each scale $(j_1,j_2)$ and location $(k_r,k_t)$, i.e.
\begin{equation}
\label{supportmr}
\mathcal{M} = \{(j_1,j_2,k_r,k_t) \big | (\mathbf{W}Y)_{j_1,j_2}[k_r,k_t] \text{ is significant}\}. 
\end{equation}
We define $\mathbf{M}$ to be the orthogonal projector onto $\mathcal{M}$, i.e. $\forall d$
\begin{equation}
(\mathbf{M}d)_{j_1,j_2}[k_r,k_t] = 
\begin{cases}
(\mathbf{W}Y)_{j_1,j_2}[k_r,k_t] & \text{if } (j_1,j_2,k_r,k_t) \in \mathcal{M}, \\
d_{j_1,j_2}[k_r,k_t] & \text{otherwise.} 
\end{cases}
\end{equation}
Our goal is to seek a solution $\widetilde{X}$ that preserves the significant structures of the original data by reproducing exactly the same wavelet coefficients as those of the input data $Y$, but only on scales and at positions where significant coefficients have been detected. Furthermore, as Poisson intensity functions are positive by nature, a positivity constraint is imposed on the solution. It is clear that there are many solutions satisfying the positivity and multiresolution support consistency requirements, e.g. $Y$ itself. Thus, our reconstruction problem based solely on these constraints is an ill-posed inverse problem that must be regularized. Typically, the solution in which we are interested must be sparse by involving the lowest budget of wavelet coefficients. Therefore, our reconstruction is formulated as a constrained sparsity-promoting minimization problem over the transform coefficients $d$
\begin{equation}
\label{minpoisson}
\min_{d} \| d \|_1 \text{ subject to } 
\begin{cases}
d_{j_1,j_2}[k_r,k_t] = (\mathbf{W}Y)_{j_1,j_2}[k_r,k_t], \forall (j_1,j_2,k_r,k_t) \in \mathcal{M}, \\ 
X \geqslant 0 .
\end{cases}
\end{equation}
and the intensity estimate $\widetilde{X}$ is reconstructed as $\widetilde{X} = \mathbf{R} \tilde{d}$, where $\tilde{d}$ is a global minimizer of Eq. \eqref{minpoisson}. We recall that $\|\cdot\|_1=\sum_{j_1,j_2,k_r,k_t}|d_{j_1,j_2}[k_r,k_t]|$ is the $\ell_1$-norm playing the role of regularization and is well-known to promote sparsity \citep{Donoho2004}. This problem can be solved efficiently using the hybrid steepest descent algorithm \citep{Yamada} \citep{Zhang}, and requires about ten iterations in practice. Transposed into our context, its main steps can be summarized as follows:

\begin{algorithm}[!h]
%\caption{MS-VSTS + IUWT Denoising}
\label{alg1}
\begin{algorithmic}[1]
\REQUIRE $\quad$ Input noisy data $Y$, a low-pass filter $h$, multiresolution support $\mathcal{M}$ from the detection step, number of iterations $N_{\max}$. \\
%\underline{\emph{\textbf{Detection}}} \\
\STATE Initialize $d^{(0)} = \mathbf{M} \mathbf{W} Y$.
\FOR{$n=1$ to $N_{\max}$}
\STATE $\bar{d}^{(n)} = \mathbf{M}d^{(n-1)}$;
\STATE $d^{(n)} = \mathbf{W} \mathbf{P}_{+}\parenth{\mathbf{R}~\mathrm{ST}_{\beta_n}(\bar{d}^{(n)})}$;
\STATE Update the step $\beta_n = (N_{\max} - n)/(N_{\max} - 1)$.
\ENDFOR
\RETURN $\widetilde{X} = \mathbf{R}d^{(N_{\max})}$.
\end{algorithmic}
\end{algorithm}
where $\mathbf{P}_+$ is the orthogonal projector onto the positive orthant, and $\mathrm{ST}_{\beta_n}$ is the entry-wise soft-thresholding operator with threshold $\beta_n$, i.e. for $x \in \mathbb{R}$, $\text{ST}_{\beta_n}(x) = \max(0,1-\beta_n/|x|)x$.

The final multichannel MS-VSTS Poisson noise removal algorithm is summarized in the following steps:
\begin{algorithm}[!h]
%\caption{MS-VSTS + IUWT Denoising}
\label{alg1}
\begin{algorithmic}[1]
\REQUIRE $\quad$ Input noisy data $Y$, a low-pass filter $h$, threshold level $\tau$. \\
%\underline{\emph{\textbf{Detection}}} \\
\STATE \emph{Multichannel spherical MS-VST:} Apply the 2D-1D MS-VSTS to $Y$ using Eqs. \eqref{detaildetailmsvst}-\eqref{approxapproxmsvst}.
\STATE \emph{Detection:} Detect the coefficients that are above $\tau$ (significant coefficients), and get the multiresolution support $\mathcal{M}$.
\STATE \emph{Reconstruction:} Apply the above algorithm with $\mathcal{M}$ to get the denoised data $\widetilde{X}$.
\end{algorithmic}
\end{algorithm}

\subsection{Experiments}

\begin{figure}[htbp]
\begin{center}
\includegraphics[width=3.8in]{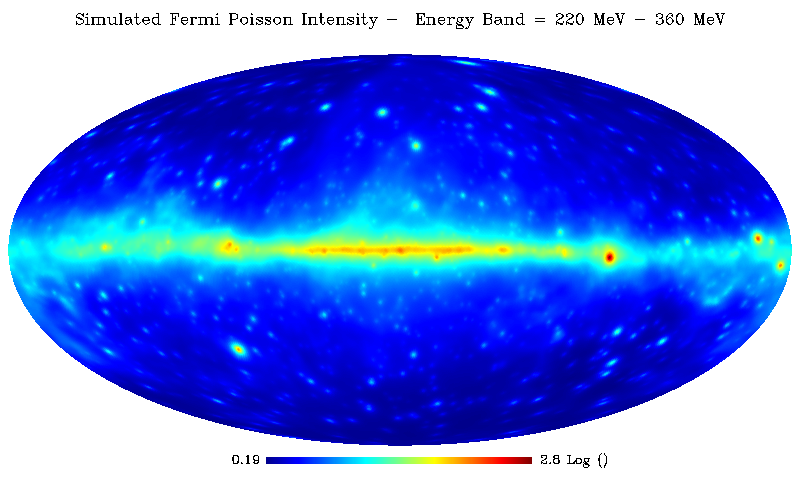} \hfill
\includegraphics[width=3.8in]{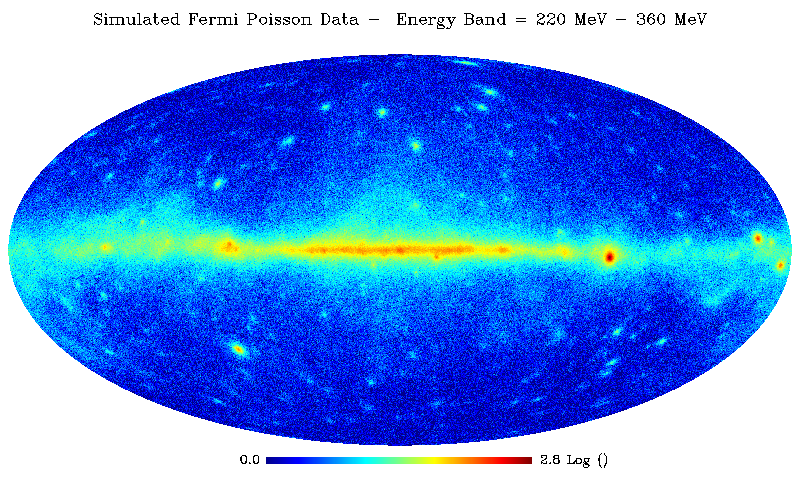} \hfill
\includegraphics[width=3.8in]{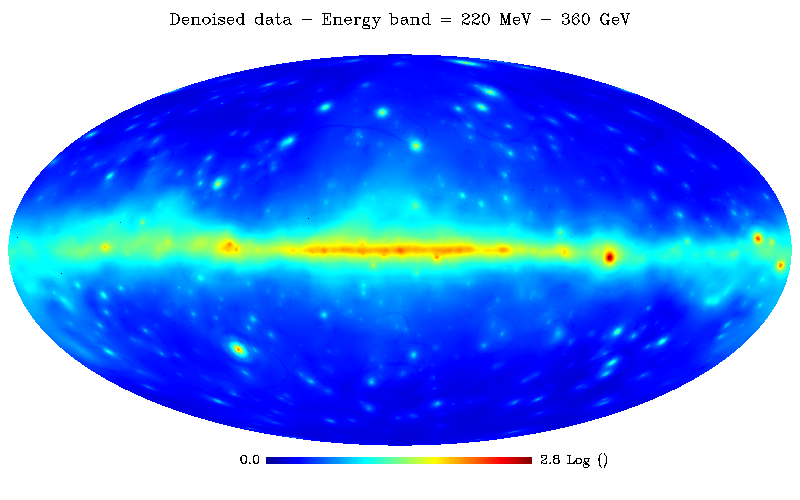} \hfill
\caption{Result of the multichannel Poisson denoising algorithm on simulated Fermi data over the energy band 220 MeV - 360 MeV. \emph{Top:} Simulated intensity skymap. \emph{Middle:} Simulated noisy skymap. \emph{Bottom:} denoised skymap.
%Result of the multichannel Poisson denoising algorithm on simulated Fermi data on two different energy bands (Top : noisy skymap. Bottom : denoised skymap).
%Energy bands: 220 MeV - 360 MeV, 589 MeV - 965 MeV.
%Energy bands: 82 MeV - 134 MeV, 220 MeV - 360 MeV, 589 MeV - 965 MeV, 1.58 GeV - 2.59 GeV, 4.24 GeV - 69.5 GeV.
Maps are on a logarithmic scale.
}
\label{denoisingmulti1}
\end{center}
\end{figure}

\begin{figure}[htbp]
\begin{center}
\includegraphics[width=3.8in]{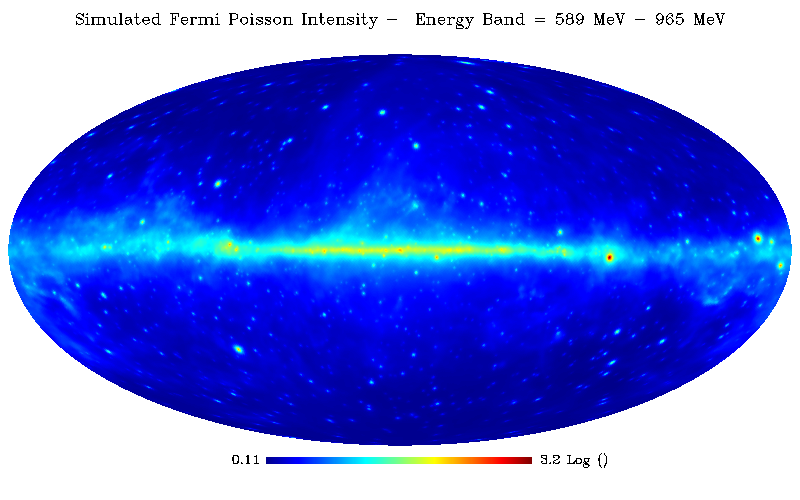} \hfill
\includegraphics[width=3.8in]{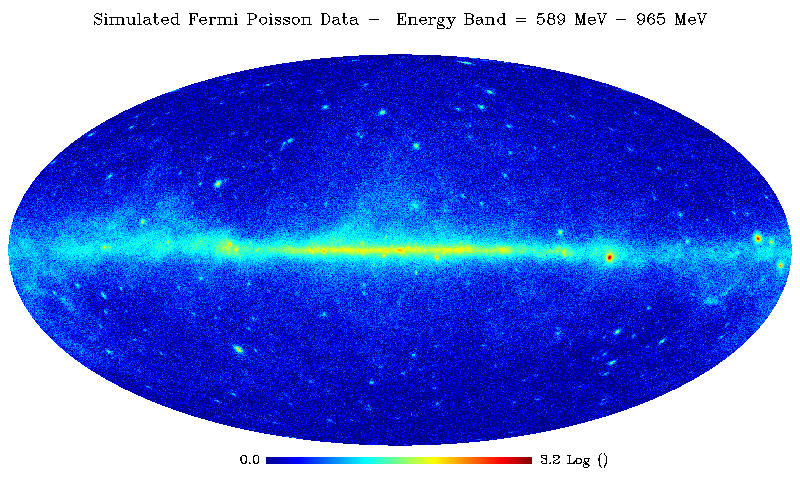} \hfill
\includegraphics[width=3.8in]{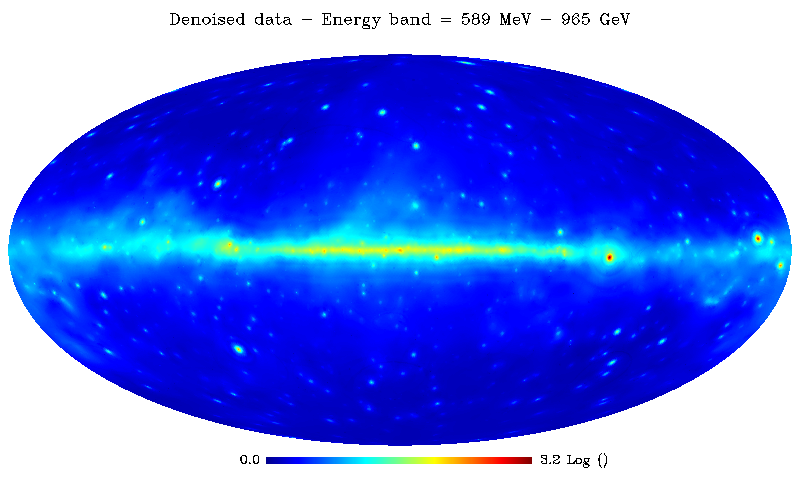} \hfill
\caption{Result of the multichannel Poisson denoising algorithm on simulated Fermi data over the energy band 589 MeV - 965 MeV. \emph{Top:} Simulated intensity skymap. \emph{Middle:} Simulated noisy skymap. \emph{Bottom:} denoised skymap.
%Result of the multichannel Poisson denoising algorithm on simulated Fermi data on two different energy bands (Top : noisy skymap. Bottom : denoised skymap).
%Energy bands: 220 MeV - 360 MeV, 589 MeV - 965 MeV.
%Energy bands: 82 MeV - 134 MeV, 220 MeV - 360 MeV, 589 MeV - 965 MeV, 1.58 GeV - 2.59 GeV, 4.24 GeV - 69.5 GeV.
Maps are on a logarithmic scale.}
\label{denoisingmulti2}
\end{center}
\end{figure}

\begin{figure}[htbp]
\begin{center}
\includegraphics[width=1in,height=0.72in]{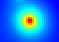}
\includegraphics[width=1in,height=0.72in]{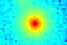}
\includegraphics[width=1in,height=0.72in]{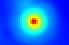} \hfill
\includegraphics[width=5in]{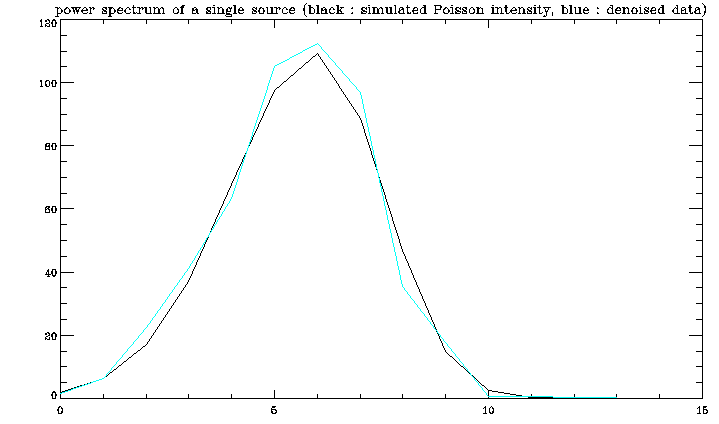}
\caption{Spectrum of a single gamma-ray point source recovered using the multichannel MS-VSTS denoising algorithm. \emph{Top:} Single gamma-ray source from simulated Fermi data integrated along the energy axis (\emph{left:} simulated source; \emph{middle:} noisy source; \emph{right:} denoised source). \emph{Bottom:} Spectrum of the center of the point source: intensity as a function of the energy band with 14 energy bands between 50 MeV and 50 GeV (\emph{black:} true simulated spectrum; \emph{cyan:} restored spectrum from our denoising algorithm.  
%(7 energy bands between 50 MeV and 1.58 GeV).
}
\label{powspec}
\end{center}
\end{figure}

%The algorithm has been applied on our simulated Fermi data set, with 14 energy bands between 50 MeV and 50 GeV. Figure \ref{denoisingmulti1} shows the result of the algorithm on 5 energy bands. %On Figure \ref{denoisingmulti2}, we compare the result of the multichannel MS-VSTS with a monochannel MS-VSTS on the data integrated along the energy axis. 
The multichannel MS-VSTS algorithm has been applied to a simulated Fermi data set, with 14 energy bands between 50 MeV and 1.58 GeV. Figures~\ref{denoisingmulti1} and \ref{denoisingmulti2} depict the denoising results for two energy bands.
The algorithm is able to recover most of the sources, even the faint ones, on each energy band. Even more importantly, the 2D-1D MS-VSTS denoising algorithm allows us to recover the spectral information for each spatial position, as can be seen from Figure~\ref{powspec}.

%\begin{figure}
%\begin{center}
%\includegraphics[width=3.9in]{datatot.png} \hfill
%\includegraphics[width=3.9in]{datatotrec.png} \hfill
%\includegraphics[width=3.9in]{solutiontot.png} \hfill
%\caption{Comparison of multichannel MS-VSTS denoising with monochannel MS-VSTS denoising. \emph{Top :} Simulated Fermi data integrated along the energy axis. \emph{Middle :} Simulated Fermi data integrated along the energy axis and denoised using monochannel MS-VSTS. \emph{Bottom :} Simulated Fermi data denoised with multichannel MS-VSTS and integrated along the energy axis.
%}
%\label{denoisingmulti2}
%\end{center}
%\end{figure}

%vue d'une face Healpix
%spectre de puissance d'un pixel

\section{Deconvolution of spherical data with Poisson noise}
\label{dec_mrs2d1d}
We now introduce a wavelet deconvolution approach for monochannel and multichannel data on the sphere with Poisson noise. The main idea underlying the method is to apply the MS-VSTS method described above. We first introduce the deconvolution problem and then describe how the MS-VSTS can be used to solve the deconvolution problem.

\subsection{Problem statement}

Many problems in signal and image processing can be cast as an inversion of a linear system

\begin{equation}
\label{linsyst}
Y = \mathbf{H} X + \varepsilon ~,
\end{equation}
where $X \in \mathcal{X}$ is the data to recover, $Y \in \mathcal{Y}$ is the degraded noisy observation, $\varepsilon$ is an additive noise, and $\mathbf{H}: \mathcal{X} \to \mathcal{Y}$ is a bounded linear operator that is typically ill-behaved because it models an acquisition process that encounters loss of information. When $ \mathbf{H}$ is the identity, it is just a denoising problem that can be treated with the previously described methods. Inverting Eq. \eqref{linsyst} is usually an ill-posed problem, which means that there is no unique and stable solution.

Our objective is to remove the effect of the instrument's PSF. In our case, $\mathbf{H}$ is the convolution by a blurring kernel (i.e. PSF) operator that causes $Y$ to lack the high frequency content of $X$. Furthermore, since the noise is Poisson, $\varepsilon$ has a variance profile $\mathbf{H} X$. The problem at hand is then a deconvolution problem in the presence of Poisson noise.

We therefore need to both regularize the \textbf{problem} and handle the Poisson statistics of the noise. To regularize this inversion problem and reduce the space of candidate solutions, one has to add some prior knowledge of the typical structure of the original data $X$. This prior information accounts for the smoothness of the solution and can range from the uniform smoothness assumption to more complex knowledge of the geometrical structures of $X$.

\begin{figure}[htbp]
\begin{center}
\includegraphics[width=6in]{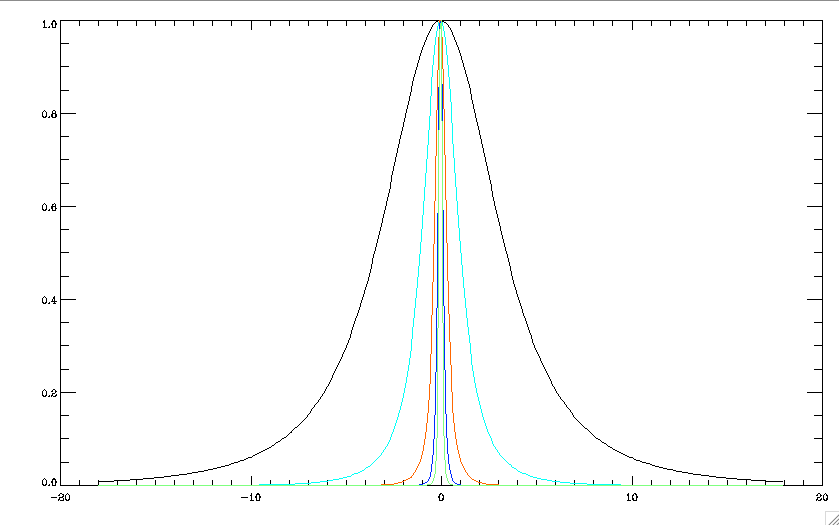} \hfill
\caption{Normalized profile of the PSF for different energy bands as a function of the angle in degree. \emph{Black:} $50$ MeV - $82$ MeV. \emph{Cyan:} $220$ MeV - $360$ MeV. \emph{Orange:} $960$ MeV - $1.6$ GeV. \emph{Blue:} $4.2$ GeV - $6.9$ GeV. \emph{Green:} $19$ GeV - $31$ GeV.}
\label{psfprofile}
\end{center}
\end{figure}

%In the case of the LAT, the point spread function depends a lot on the energy, from about $3.5°$ at $100$ MeV to better than $0.1°$ at $10$ GeV and above.
In our LAT realistic simulations, the PSF width depends strongly on the energy, from $6.9°$ at $50$ MeV to better than $0.1°$ at $10$ GeV and above. Figure~\ref{psfprofile} shows the normalized profiles of the PSF for different energy bands.

%exemples de PSF

\subsection{Monochannel Deconvolution}

We first consider the single-channel case. In the literature, several algorithms have been proposed to perform image deconvolution on a cartesian grid. The Richardson-Lucy algorithm is certainly the most famous in astrophysics. In this paper, we propose a regularized Richardson-Lucy algorithm to deconvolve data on the sphere data.

The Richardson-Lucy algorithm originates from a fixed-point equation obtained by maximizing the Poisson likelihood with respect to $X$ while preserving positivity. This entails a multiplicative update rule, starting at $n=0$ and $X^{(0)} = 1$ and iterating
\begin{equation}
\label{JVCmethod}
X^{(n+1)} = X^{(n)} \otimes \parenth{ \mathbf{H}^{T} (Y \odiv {\mathbf{H} X^{(n)}})},
\end{equation}
where $\otimes$ (respectively $\odiv$) represents the element-wise multiplication (respectively division) between two vectors, and $\mathbf{H}^{T}$ is the transpose of $\mathbf{H}$ whose action on an image consists in convolving it with the time-reversed version of the PSF associated with $\mathbf{H}$. However, it is well-known that owing to the lack of regularization, the Richardson-Lucy algorithm tends to amplify the noise after a few iterations.

We define $R^{(n)}$ as the residual at iteration $n$
\begin{equation}
\label{residual}
R^{(n)} = Y- \mathbf{H} X^{(n)},
\end{equation}
where $R^{(n)}$ can be written as the sum of its IUWTS detail subband $\{d_j\}_{1 \leqslant j \leqslant J}$ and the last approximation subband $a_J$, that is,
\begin{equation}
\label{waveletres}
R^{(n)}[k_r] = a_J [k_r]  + \sum_{j=1}^{J} d_j [k_r], \quad \forall k_r.
\end{equation}
The wavelet transform provides a means of extracting only the significant structures from the residual at each iteration. With increasing number of iterations, a large part of the residual becomes statistically insignificant. The regularized significant residual is then, for a location $k_r$
\begin{equation}
\label{signifgauss}
\widebar{R}^{(n)}[k_r] = a_J [k_r]  + \sum_{(j,k_r) \in \mathcal{M}} d_j [k_r] ~,
\end{equation}
where $\mathcal{M}$ is the multiresolution support defined in a similar way to Eq. \eqref{supportmr}. The regularized Richardson-Lucy scheme then becomes 
\begin{equation}
\label{regularizedJVC}
X^{(n+1)} = \mathbf{P}_+\parenth{X^{(n)} \otimes \parenth{ \mathbf{H}^{T} \parenth{({\mathbf{H} X^{(n)}}+\widebar{R}^{(n)}) \odiv {\mathbf{H} X^{(n)}}}}} ~.
\end{equation}
This algorithm is similar to \citet{starck:mur95_2}, except that the {\textit{\`a trous}} wavelet transform is replaced by the undecimated wavelet transform. In the next subsection, we describe how the same algorithm can be extended to the multi-channel case.

\subsection{Multichannel deconvolution}
As the PSF is channel-dependent, the convolution observation model is now
\[
Y[\cdot,k_t] = \mathbf{H}_{k_t} X[\cdot,k_t] + \varepsilon[\cdot,k_t]
\]
in each channel $k_t$, where $\mathbf{H}_{k_t}$ is the (spatial) convolution operator in channel $k_t$ with known PSF.

%In this problem, the data can be viewed as a matrix $Y=(Y_t)_t$, where, for each channel $t$, $Y_t = (y_{r,t})_{r}$ is the vector corresponding to the spherical data at channel $t$, where $r$ is the index corresponding to the pixel. Each channel is convolved by a known blurring kernel $\mathbf{ H}_t$ : $Y_t = \mathbf{ H}_t \star X_t + \mathcal{N}(X_t)$.

The same recipe as in the monochannel case applies with the notable difference that the spherical 2D-1D MS-VSTS is used instead of its monochannel counterpart. The multichannel multiresolution support $\mathcal{M}$ is obtained after thresholding these coefficients. 

We now define $\mathbf{H}$ to be the multichannel convolution\footnote{Strictly speaking, this is a slight abuse of terminology since the kernel is not channel-invariant.} operator, which acts on a 2D-1D multichannel spherical data set $X$ by applying $\mathbf{H}_{t}$ to each channel $X[\cdot,k_t]$ independently\footnote{If $X$ were to be vectorized by stacking the channels in a long column vector, $\mathbf{H}$ would be a block-diagonal matrix whose blocks are the circulant matrices $\mathbf{H}_{k_t}$.}. The regularized multichannel Richardson-Lucy scheme is then
%We denote $\mathcal{H}$ the convolution by channel operator: $\mathcal{H} \mathbf{ X}$ means that each channel $X_t$ is convolved by the convolution kernel $\mathbf{ H}_t$. $\mathcal{H}^T \mathbf{ X}$ means that each channel $X_t$ is convolved by the transposed convolution kernel $\mathbf{ H}^T_t$. 
\begin{equation}
\label{regularizedmcdeconv}
X^{(n+1)} = \mathbf{P}_+\parenth{X^{(n)} \otimes \parenth{ \mathbf{H}^{T} \parenth{({\mathbf{H} X^{(n)}}+\widebar{R}^{(n)}) \odiv {\mathbf{H} X^{(n)}}}}} ~,
\end{equation}
where $\widebar{R}^{(n)}$ is the regularized (significant) residual
\begin{equation}
\begin{split}
\label{signifresmc}
\widebar{R}^{(n)}[k_r,k_t] = a_{J_1,J_2} [k_r,k_t] +  \sum_{(j_1,j_2,k_r,k_t) \in \mathcal{M}} w_{j_1,j_2}[k_r,k_t] ~.
\end{split}
\end{equation}

%The reconstruction algorithm is:

%\begin{algorithm}[!h]
%%\caption{MS-VSTS + IUWT Denoising}
%\label{alg1}
%\begin{algorithmic}[1]
%\REQUIRE $\quad$ Input noisy data $Y$, a low-pass filter $h$, multiresolution support $\mathcal{M}$ from the detection step, number of iterations $N_{\max}$ \\
%%\underline{\emph{\textbf{Detection}}} \\
%\STATE Initialize $X^{(0)} = \mathcal{M} \mathbf{W} Y = \mathcal{M}w_{Y}$,
%\FOR{$n=1$ to $N_{\max}$}
%\STATE $\widetilde{d} = \mathcal{M}w_Y + (1-\mathcal{M})\mathbf{W}\mathcal{H}X^{(n-1)}$,
%\STATE $X^{(n)} = P_{+}(\mathcal{H}^{T}\mathbf{R}\text{ST}_{\beta_n}[\widetilde{d}])$,
%\STATE Update the step $\beta_n = (N_{\max} - n)/(N_{\max} - 1)$
%\ENDFOR

%\end{algorithmic}
%\end{algorithm}

\subsection{Experiments}

\begin{figure}[htbp]
\begin{center}
\includegraphics[width=3.8in]{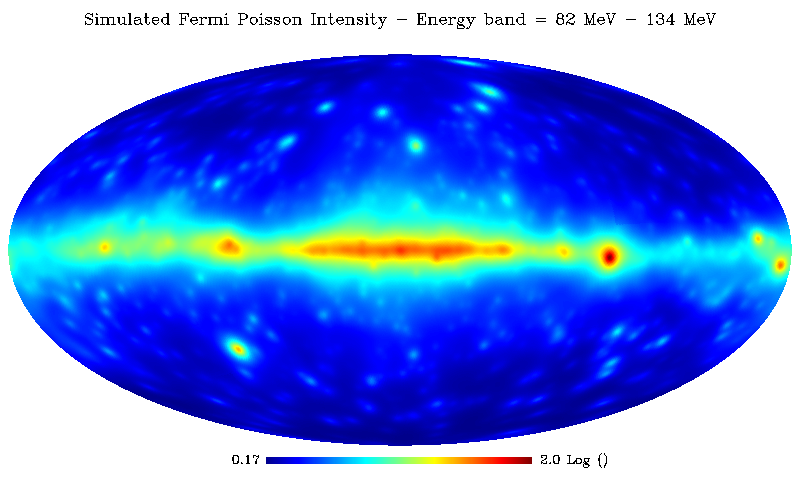} \hfill
\includegraphics[width=3.8in]{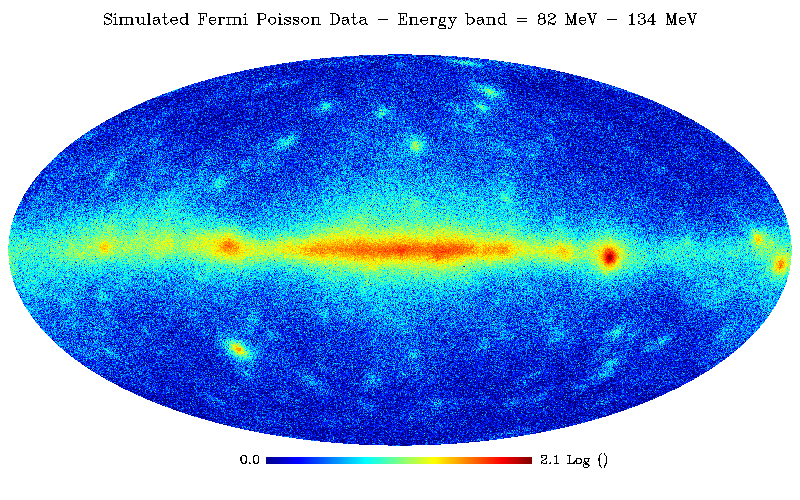} \hfill
\includegraphics[width=3.8in]{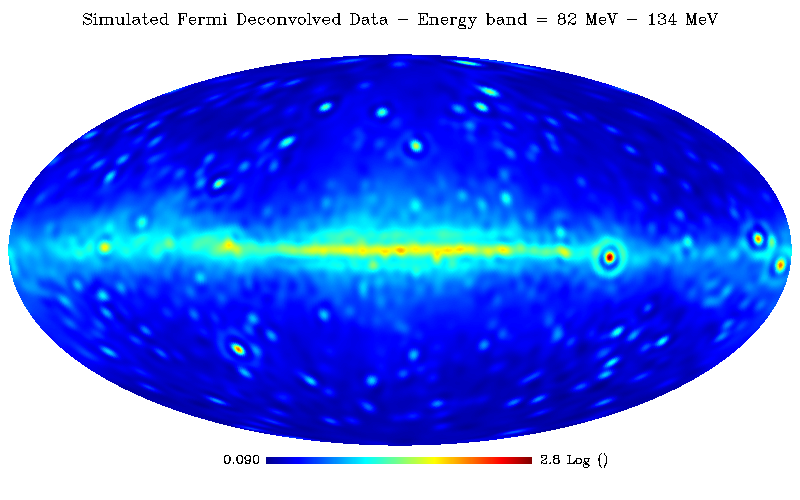}
\caption{Result of the multichannel deconvolution algorithm for different energy bands. \emph{Top:} Simulated (blurred) intensity skymap. \emph{Middle:} Blurred and noisy skymap. \emph{Bottom:} Deconvolved skymap.
%Energy bands : 50 MeV - 82 MeV, 82 MeV - 134 MeV, 134 MeV - 220 MeV, 220 MeV - 360 MeV.
Energy band : 82 MeV - 134 MeV. Maps are on a logarithmic scale.
}
\label{deconv1}
\end{center}
\end{figure}

\begin{figure}[htbp]
\begin{center}
\includegraphics[width=3.8in]{plot_mc_intensityband3.png} \hfill
\includegraphics[width=3.8in]{plot_mc_simupoissonband3.png} \hfill
\includegraphics[width=3.8in]{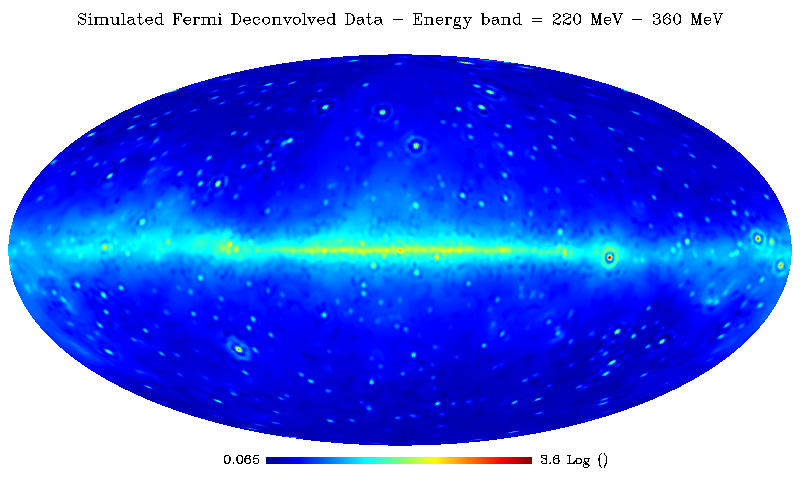}
\caption{Result of the multichannel deconvolution algorithm for different energy bands. \emph{Top:} Simulated (blurred) intensity skymap. \emph{Middle:} Blurred and noisy skymap. \emph{Bottom:} Deconvolved skymap.
%Energy bands : 50 MeV - 82 MeV, 82 MeV - 134 MeV, 134 MeV - 220 MeV, 220 MeV - 360 MeV.
Energy band : 220 MeV - 360 MeV. Maps are on a logarithmic scale.
}
\label{deconv2}
\end{center}
\end{figure}

\begin{figure}[htbp]
\begin{center}
\includegraphics[width=3.8in]{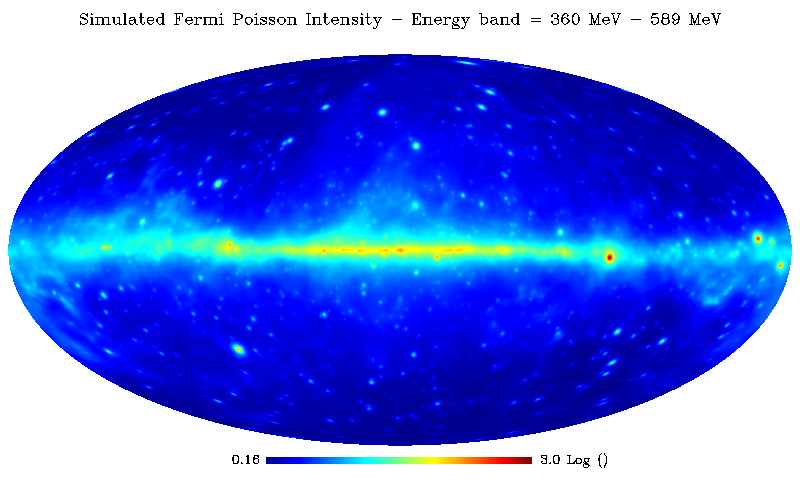} \hfill
\includegraphics[width=3.8in]{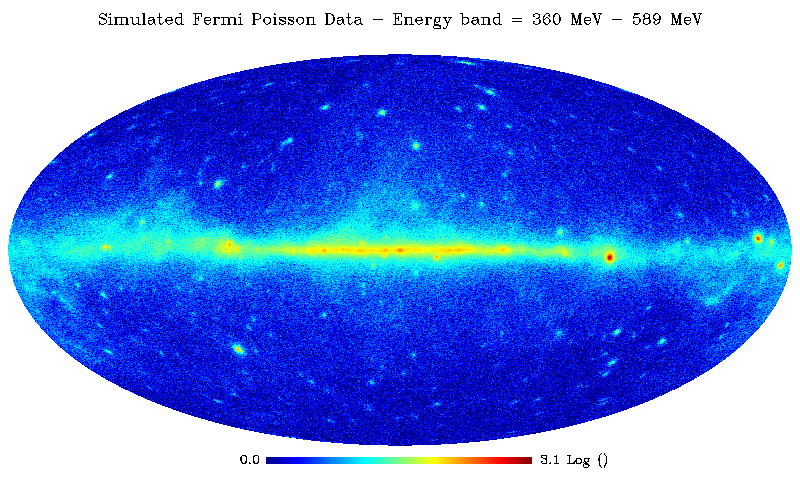} \hfill
\includegraphics[width=3.8in]{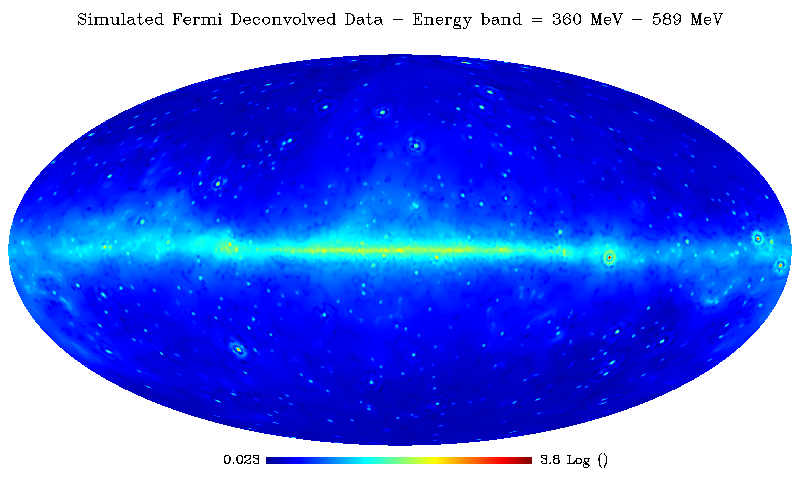}
\caption{Result of the multichannel deconvolution algorithm for different energy bands. \emph{Top:} Simulated (blurred) intensity skymap. \emph{Middle:} Blurred and noisy skymap. \emph{Bottom:} Deconvolved skymap.
%Energy bands : 50 MeV - 82 MeV, 82 MeV - 134 MeV, 134 MeV - 220 MeV, 220 MeV - 360 MeV.
Energy band : 360 MeV - 589 MeV. Maps are on a logarithmic scale.
}
\label{deconv3}
\end{center}
\end{figure}

\begin{figure}[htbp]
\begin{center}
\includegraphics[width=3.8in]{plot_mc_intensityband5.png} \hfill
\includegraphics[width=3.8in]{plot_mc_simupoissonband5.png} \hfill
\includegraphics[width=3.8in]{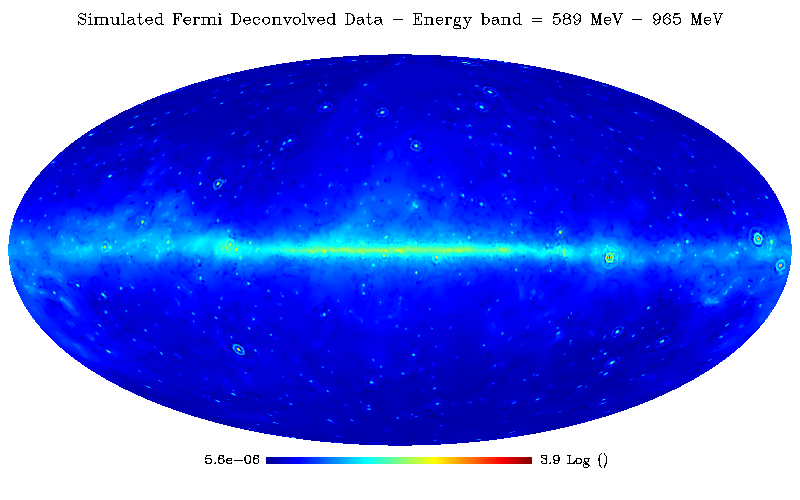}
\caption{Result of the multichannel deconvolution algorithm for different energy bands. \emph{Top:} Simulated (blurred) intensity skymap. \emph{Middle:} Blurred and noisy skymap. \emph{Bottom:} Deconvolved skymap.
%Energy bands : 50 MeV - 82 MeV, 82 MeV - 134 MeV, 134 MeV - 220 MeV, 220 MeV - 360 MeV.
Energy band : 589 MeV - 965 MeV. Maps are on a logarithmic scale.
}
\label{deconv4}
\end{center}
\end{figure}

\begin{figure}[htbp]
\begin{center}
\includegraphics[width=1in,height=0.72in]{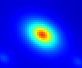}
\includegraphics[width=1in,height=0.72in]{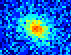}
\includegraphics[width=1in,height=0.72in]{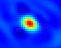} \hfill
\includegraphics[width=5in]{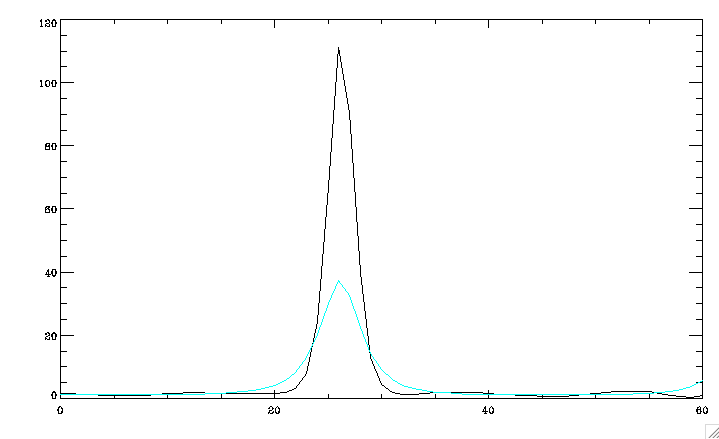}
\caption{Profile of a single gamma-ray point source recovered using the multichannel MS-VSTS deconvolution algorithm. \emph{Top:} Single gamma-ray point source on simulated (blurred) Fermi data (energy band: 360 MeV - 589 MeV) (\emph{left:} simulated blurred source; \emph{middle:} blurred noisy source; \emph{right:} deconvolved source). \emph{Bottom:} Profile of the point source (\emph{cyan:} simulated spectrum; \emph{black:} restored spectrum from the deconvolved source.
}
\label{deconv5}
\end{center}
\end{figure}

\begin{figure}[htbp]
\begin{center}
\includegraphics[width=2.9in]{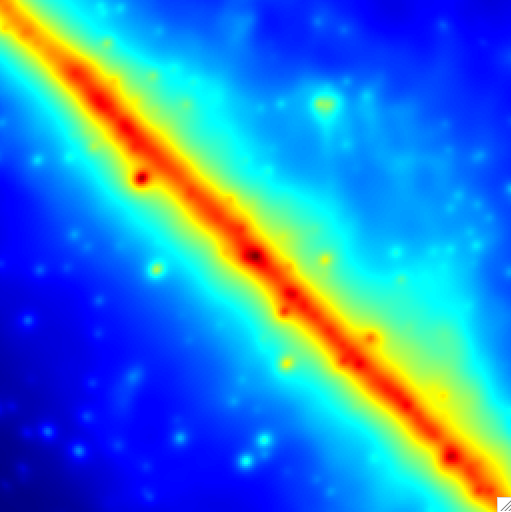} \hfill
\includegraphics[width=2.9in]{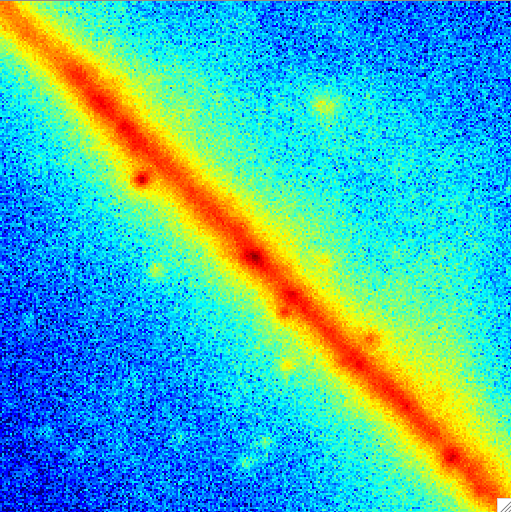} \hfill
\includegraphics[width=2.9in]{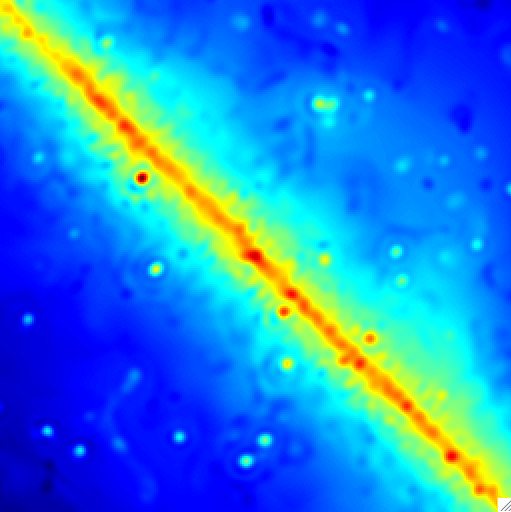}
\caption{View on a single HEALPix face. Result of an application of the deconvolution algorithm to the Galactic plane. \emph{Top Left:} Simulated Fermi Poisson intensity. \emph{Top Right:} Simulated Fermi noisy data. \emph{Bottom:} Fermi data deconvolved with multichannel MS-VSTS. 
Energy band: 360 MeV - 589 MeV.
Pictures are on a logarithmic scale.
}
\label{deconv6}
\end{center}
\end{figure}

%The algorithm is applied on the 7 first energy bands of our simulated Fermi data set ($50 MeV$ to $1.58 GeV$).
The algorithm was applied to the seven energy bands ($50$ MeV-$1.58$ GeV) of our simulated Fermi data set. Figures~\ref{deconv1} to \ref{deconv4} display the deconvolution results for four energy bands.  %Figure \ref{deconv3} compares the result of the multichannel MS-VSTS deconvolution with the result of the multichannel MS-VSTS denoising on data integrated along the energy axis.
Figure~\ref{deconv5} shows the performance of the multichannel MS-VSTS deconvolution algorithm for a single point source. The deconvolution not only effectively removes the blur and recovers sharply localized point sources, but also allows us to restore all of the spectral information. To get a better visual impression of the performance of the deconvolution algorithm, Figure~\ref{deconv6} depicts the result of the algorithm on a single HEALPix face covering the Galactic plane. We find that the deconvolution is remarkably effective : our MS-VSTS multichannel deconvolution algorithm manages to remove a large part of the blur introduced by the PSF.

%\begin{figure}
%\begin{center}
%\includegraphics[width=2.9in]{datatotnoisyface.png} \hfill
%\includegraphics[width=2.9in]{datatotdenoisedface.png} \hfill
%\includegraphics[width=2.9in]{datatotdeconvface.png}
%\caption{View on a single HEALPix face. Comparison of the multichannel denoising algorithm with the multichannel deconvolution algorithm on the galactic plan. \emph{Top Left:} Simulated Fermi data integrated along the energy axis. \emph{Top Right:} Simulated Fermi data denoised with multichannel MS-VSTS and integrated along the energy axis. \emph{Bottom:} Fermi data deconvolved with multichannel MS-VSTS and integrated along the energy axis. 
%%Energy bands : 360 MeV - 589 MeV, 589 MeV - 965 MeV, 965 MeV - 1.58 GeV.
%Pictures are in logarithmic scale.
%}
%\label{deconvtot}
%\end{center}
%\end{figure}

\section*{Software}
The software related to this paper, {\bf MRS/Poisson}, and its full documentation will be included in the next version of ISAP (Interactive Sparse astronomical data Analysis Packages) via the ISAP web site.\footnote{http://jstarck.free.fr/isap.html}
%The software related to this paper, {\bf MRS/Poisson}, and its full documentation will be included in the next version of ISAP (Interactive Sparse astronomical data Analysis Packages) via the web site:\\ \\
%{\centerline{\texttt{http://jstarck.free.fr/isap.html}}}\\

\section{Conclusion}
\label{sec:conclusion}

This paper extends the MS-VSTS framework to deal with monochannel deconvolution, multichannel denoising and multichannel deconvolution. Unlike the monochannel MS-VSTS, the multichannel MS-VSTS fully exploits the information in the 2D-1D  data set and allows us to recover the spectral information on the sources. As the PSF strongly depends on the energy, it is very important to have a multichannel method for deconvolution. Multichannel deconvolution using MS-VSTS removes a large part of the PSF blur and significantly improves the sharpness of the spatial localization of point sources.

\begin{acknowledgements}
Some of the results in this paper have been derived using the Healpix \citep{Gorski} and  the MR/S software \citep{starck2006}. This work was supported by the French National Agency for Research (ANR-08-EMER-009-01) and the European Research Council grant SparseAstro (ERC-228261). 
\end{acknowledgements}

\bibliographystyle{aa} % style aa.bst
\bibliography{articlemc.bib} % your references Yourfile.bib

\end{document}